\documentstyle[12pt,times,psfig]{article}
\newcommand{\beq}{\begin{equation}}
\newcommand{\eeq}{\end{equation}}
\newcommand{\bea}{\begin{eqnarray}}
\newcommand{\eea}{\end{eqnarray}}

\begin{document}
\noindent{
{\large\bf 
LANL Report LA-UR-98-6000 (1998)\\
}
}
%{\small
Talk given at the {\em Fourth Workshop on Simulating Accelerator Radiation
Environments (SARE4)}, \\
Knoxville, Tennessee, September 14-16, 1998
%}

\begin{center}
{\Large \bf 
Production and Validation of Isotope Production
Cross Section Libraries for Neutrons and Protons to 1.7 GeV
}\\
\vspace*{0.3cm}
{\bf S.~G.~Mashnik, A. J. Sierk, K. A. Van Riper,$^*$ and W. B. Wilson}\\
\vspace{0.3cm}
{\it T-2, Theoretical Division, Los Alamos National Laboratory,
Los Alamos, NM 87545}\\
{\it $^*$White Rock Science, PO Box 4729,
Los Alamos, NM 87545}\\
\end{center}
\begin{abstract}
For validation and development of codes and for modeling isotope production in 
high power accelerators and APT Materials studies,
% and other applications,  
we have produced experimental, calculated, and evaluated activation libraries 
for interaction of nucleons with nuclides covering about a third of all natural elements.
For targets considered here, our compilation of experimental data
is the most complete we are aware of, since it contains 
all data available on the Web, in journal papers, laboratory reports,
theses, and books,
% we were able to find,
as well as all data included in the large
compilation by Sobolevsky with co-authors (NUCLEX) published recently 
by Springer-Verlag in 4 volumes.
%, as well as about half of new data not covered by NUCLEX. 
Our evaluated library was produced using all
available experimental cross sections together with calculations
by the CEM95, LAHET, and HMS-ALICE codes and with the European Activation File
EAF-97 and LANL Update II of the ECNAF Neutron Activation Cross-Section
Library. 
\end{abstract}

\begin{center}
{\large 1. Introduction} \\
\end{center}

Data on isotope production yields from various reactions are 
necessary to validate and develop models of nuclear reactions and are
a decisive input for many applications, e.g., for
accelerator transmutation of waste (ATW), 
% for elimination of long-lived radioactive wastes with a spallation source,
accelerator-based conversion (ABC), 
% aimed to complete the destruction of weapon plutonium,
accelerator-driven energy production (ADEP), 
% which proposes to derive fission energy from thorium with 
% concurrent destruction of the long-lived waste
% and without the production of weapon-usable material, 
accelerator production of tritium (APT),
for the optimization of commercial production of radioisotopes  
used in medicine, mining, and industry, for solving problems of 
radiation protection of cosmonauts, aviators, workers at nuclear
facilities, and for modeling radiation damage to computer chips, etc.
 (see details and references  in \cite{report97}).  
Also,
residual product nuclide yields in thin targets irradiated by medium-
and high-energy projectiles are extensively used in cosmochemistry and
cosmophysics to interpret the production of cosmogenic nuclides in 
meteorites by primary galactic particles.

Ideally, it would be desirable to have a universal library that includes data for 
all nuclides, projectiles, and incident energies. At present,
neither the measurements nor the available codes permit one to create a
reliable library covering all data needed at intermediate energies.
First, experiments are costly and it is impossible to measure all
data, in principle. Second, predictions by the best of available
models, codes, and phenomenological systematics may differ for
yields of certain isotopes at energies above 100 MeV by a factor of
100 or more, so that current models
have to be further developed before they become reliable predictive 
tools (see, e.g., \cite{report97}). Construction of a universal
comprehensive library would be very time consuming and costly, and is beyond
the scope of the present work. Instead, we have collected experimental data, 
performed calculations with the most reliable codes, and evaluated cross
sections for nucleon-nucleus interactions at intermediate energies for
a number of targets of interest. Our progress in this activity is
briefly described in this paper.

\begin{center}
{\large 2. Experimental Data Library} \\
\end{center}

We compile nucleon-induced isotope production experimental
cross sections for two reasons. First, for validation \cite{report97}
and further development \cite{cem98}
of the Cascade-Exciton Model (CEM) \cite{cem} code 
CEM95 \cite{cem95} and to investigate the applicability of the
recent ALICE code with the new Hybrid
Monte Carlo Simulation model (HMS-ALICE) \cite{alice96}
to produce activation libraries up to 150 MeV \cite{mark,act150}.
Second, we need these data to evaluate reliable libraries 
for our medical isotope production \cite{medical97,medical98}  
and material studies \cite{mike98}
at APT. For these tasks, we need data only up to several GeV. But
realizing that such a compilation will be useful in the future for
other problems, we do not limit ourselves to these energies, 
but compile all available data at any energy above several MeV.
Since there are very few data on neutron-induced
isotope production cross sections at energies above 100 MeV, 
we focus mainly on compiling proton-induced reaction data.

Many efforts have been previously made to compile 
experimental production yields from proton-induced reactions
at intermediate energies.  To the best of our knowledge,
the most complete compilation was performed by Sobolevsky and
co-authors and was published recently by Springer-Verlag in eight
separate subvolumes \cite{nuclex}. 
Sobolevsky and co-authors have performed a major work and compiled
all data available to them for target elements from Helium to 
transuranics for the entire energy range from thresholds up
to the highest energy measured.
For proton-induced reactions, this compilation contains about 37,000
data points  published in the first four Subvolumes {\bf I/13a-d} \cite{nuclex}
(the following Subvolumes {\bf I/13e-h} concern pion, deutron, triton, $^3$He,
and alpha induced reactions). 
This valuable compilation is also currently available
in an electronic version as an IBM PC code named NUCLEX \cite{nuclex}.
Unfortunately, since this compilation is very expensive (Springer Verlag
sells a single new subvolume for \$1647.00 or \$2020.00; interested buyers 
may find information at: 
{\bf http://www.springer-ny.com/catalog/np/nov96np/DATA/3-540-61045-6.html})
it is not easily available to individual researchers or to small libraries.
Of more immediate concern is the fact that NUCLEX does not contain a large volume of
data obtained during recent years, especially for proton-induced reactions. 

Due to the increasing interest in intermediate-energy data
for ATW, ABC, ADEP, APT, astrophysics, and other applications,  
precise and voluminous measurements of proton-induced spallation
cross sections have been performed recently, and are presently in progress,
by R. Michel et al. from Hannover University \cite{michel},
Yu. E. Titarenko et al. at ITEP, Moscow \cite{titarenko},
Yu. V. Aleksandrov et al. at JINR, Dubna \cite{al94},
B. N. Belyaev et al. at B. P. Konstantinov St. Petersburg Institute of
Nuclear Physics \cite{bel92},
N. I. Venikov et al. at Kurchatov Institute, Moscow \cite{ven93}
A. S. Danagulyan et al. at JINR, Dubna \cite{dan97},
H. Vonach et al. at LANL, Los Alamos \cite{von97},
S. Sudar and S. M. Qaim at KFA, J\"ulich \cite{su94},
D. W. Bardayan et al. at LBNL, Berkeley \cite{ba97},
J. M. Sisterson et al. at TRIUMF and other accelerators \cite{si97}, etc.
Finally, we note another, ``new" type of nuclear reaction
intensively studied in recent years, which provides irreplacable
data for our needs. These are from reactions using reverse kinematics,
when relativistic ions interact with hydrogen targets and they often
provide the only way to obtain reliable data for interaction
of intermediate energy 
protons with separate isotopes of an element with a complex
natural isotopic composition. Much data from this type of reaction
have been recently obtained, e.g.,
by W. R. Webber et al. at the LBL Bevalac \cite{web90,chen97} and
L. Tassan-Got et al. at GSI, Darmstadt \cite{ta98}.
These new data, as well as a number of other new and old measurements
have not been covered by NUCLEX. Therefore, we do not confine ourselves
solely to NUCLEX as a source of experimental cross sections; instead, we 
compile all available data for the targets in which we are interested,
searching first the World Wide Web, then any other sources available
to us, including the compilation from NUCLEX.

Table 1 shows the whole list of isotope production cross sections
we have so far included in our experimental library file.
Actually, our experimental library consists at present of 32 files,
a separate file for each element, stored in a simple and easy to read
format, and a README file with the description of 

\newpage
\begin{center}
{\bf Table 1}\\
Experimental Data Library for Proton-Induced Isotope Production
\end{center}

\begin{center}
\begin{tabular}{ccccccc}
\hline 
\hline 
Target Z & Target \# & Target & No. of  & The same,  &
No. of data  & The same,  \\
 &  &  & reactions & in NUCLEX \cite{nuclex}&
points & in NUCLEX \cite{nuclex} \\
\hline 
\hline 
6 &    &           &     &    & 652 & 487 \\
  & 1  & $^{13}$C  & 1   & 1  &     &     \\
  & 2  & $^{12}$C  & 7   & 0  &     &     \\
  & 3  & $^{nat}$C & 13  & 8  &     &     \\
\hline 
7 &    &           &     &    & 493 & 419 \\
  & 4  & $^{14}$N  & 11  & 0  &     &     \\
  & 5  & $^{13}$N  & 5   & 5  &     &     \\
  & 6  & $^{nat}$N & 9   & 8  &     &     \\
\hline 
8 &    &           &     &    & 400 & 347 \\
  & 7  & $^{18}$O  & 5   & 5  &     &     \\
  & 8  & $^{16}$O  & 14  & 0  &     &     \\
  & 9  & $^{nat}$O & 16  & 11 &     &     \\
\hline 
9 &    &           &     &    & 153 & 136 \\
  & 10 & $^{19}$F  & 9   & 9  &     &     \\
\hline 
10&    &           &     &    & 28  & 7   \\
  & 11 & $^{22}$Ne &  1  & 1  &     &     \\
  & 12 & $^{20}$Ne & 21  & 0  &     &     \\
  & 13 & $^{nat}$Ne & 4   & 4  &     &     \\
\hline 
11&    &           &     &    & 96  & 91  \\
  & 14 & $^{23}$Na & 8   & 8  &     &     \\
\hline 
12&    &           &     &    & 596 & 497 \\
  & 15 & $^{26}$Mg & 3   & 3  &     &     \\
  & 16 & $^{25}$Mg & 3   & 3  &     &     \\
  & 17 & $^{24}$Mg & 26  & 3  &     &     \\
  & 18 & $^{nat}$Mg& 17  & 14 &     &     \\
\hline 
13&    &           &     &    & 1170& 816 \\
  & 19 & $^{27}$Al & 45  & 23 &     &     \\
\hline 
15&    &           &     &    & 42  & 37  \\
  & 20 & $^{31}$P  & 5   & 4  &     &     \\
\hline 
16&    &           &     &    & 93  & 49  \\
  & 21 & $^{34}$S  & 1   & 1  &     &     \\
  & 22 & $^{32}$S  & 33  & 0  &     &     \\
  & 23 & $^{nat}$S & 10  & 9  &     &     \\
\hline 
17&    &            &     &    & 25  & 25  \\
  & 24 & $^{nat}$Cl & 4   & 4  &     &     \\
\hline 
18&    &            &     &    & 102 & 66  \\
  & 25 & $^{40}$Ar  & 36  & 0  &     &     \\
  & 26 & $^{nat}$Ar & 17  & 17 &     &     \\
\hline 
19&    &            &     &    & 20  & 20  \\
  & 27 & $^{nat}$K  & 3   & 3  &     &     \\
\hline 
\hline 
\end{tabular}
\end{center}

\begin{center}
{\bf Table 1 (continued)}\\
\end{center}

\begin{center}
\begin{tabular}{ccccccc}
\hline 
\hline 
Target Z & Target \# & Target & No. of  & The same,  &
No. of data  & The same,  \\
 &  &  & reactions & in NUCLEX \cite{nuclex} &
points & in NUCLEX \cite{nuclex} \\
\hline 
\hline 
20&    &            &     &    & 660 & 258 \\
  & 28 & $^{48}$Ca  & 1   & 1  &     &     \\
  & 29 & $^{44}$Ca  & 4   & 4  &     &     \\
  & 30 & $^{43}$Ca  & 2   & 2  &     &     \\
  & 31 & $^{42}$Ca  & 1   & 1  &     &     \\
  & 32 & $^{40}$Ca  & 57  & 0  &     &     \\
  & 33 & $^{nat}$Ca & 26  & 12 &     &     \\
\hline 
26&    &            &     &    & 2638& 1189\\
  & 34 & $^{58}$Fe  & 4   & 4  &     &     \\
  & 35 & $^{57}$Fe  & 5   & 5  &     &     \\
  & 36 & $^{56}$Fe  & 99  & 7  &     &     \\
  & 37 & $^{44}$Fe  & 2   & 2  &     &     \\
  & 38 & $^{nat}$Fe & 92  & 58 &     &     \\
\hline 
27&    &            &     &    & 2213& 871 \\
  & 39 & $^{59}$Co  & 54  & 48 &     &     \\
\hline 
30&    &           &     &    & 997 & 996 \\
  & 40 & $^{70}$Zn & 2   & 2  &     &     \\
  & 41 & $^{68}$Zn & 6   & 6  &     &     \\
  & 42 & $^{67}$Zn & 3   & 3  &     &     \\
  & 43 & $^{66}$Zn & 3   & 3  &     &     \\
  & 44 & $^{64}$Zn & 6   & 6  &     &     \\
  & 45 & $^{nat}$Zn& 43  & 42 &     &     \\
\hline 
31&    &           &     &    & 356 & 356 \\
  & 46 & $^{71}$Ga & 12  & 12 &     &     \\
  & 47 & $^{69}$Ga & 13  & 13 &     &     \\
  & 48 & $^{nat}$Ga& 6   & 6  &     &     \\
\hline 
32&    &           &     &    & 651 & 651 \\
  & 49 & $^{76}$Ge & 31  & 31 &     &     \\
  & 50 & $^{74}$Ge & 3   & 3  &     &     \\
  & 51 & $^{73}$Ge & 2   & 2  &     &     \\
  & 52 & $^{72}$Ge & 6   & 6  &     &     \\
  & 53 & $^{70}$Ge & 29  & 29 &     &     \\
  & 54 & $^{nat}$Ge& 15  & 15 &     &     \\
\hline 
33&    &           &     &    & 198 & 180 \\
  & 55 & $^{75}$As & 62  & 59 &     &     \\
\hline 
39&    &           &     &    & 1509&  614\\
  & 56 & $^{89}$Y  & 65  & 50 &     &     \\
\hline 
40&    &           &     &    & 2351&  812\\
  & 57 & $^{96}$Zr & 14  & 14 &     &     \\
  & 58 & $^{94}$Zr & 30  & 30 &     &     \\
  & 59 & $^{92}$Zr &  4  & 4  &     &     \\
\hline 
\hline 
\end{tabular}
\end{center}

\begin{center}
{\bf Table 1 (continued)}\\
\end{center}

\begin{center}
\begin{tabular}{ccccccc}
\hline 
\hline 
Target Z & Target \# & Target & No. of  & The same,  &
No. of data  & The same,  \\
 &  &  & reactions & in NUCLEX \cite{nuclex} &
points & in NUCLEX \cite{nuclex} \\
\hline 
\hline 
40&    &           &     &    & 2351&  812\\
  & 60 & $^{91}$Zr & 44  & 42 &     &     \\
  & 61 & $^{90}$Zr & 40  & 40 &     &     \\
  & 62 & $^{nat}$Zr& 73  & 34 &     &     \\
\hline 
41&    &           &     &    & 898 &  205\\
  & 63 & $^{93}$Nb & 70  & 61 &     &     \\
\hline 
42&    &           &     &    & 1432& 1339\\
  & 64 & $^{100}$Mo& 8   & 8  &     &     \\
  & 65 & $^{98}$Mo & 7   & 7  &     &     \\
  & 66 & $^{97}$Mo & 5   & 5  &     &     \\
  & 67 & $^{96}$Mo & 25  & 25 &     &     \\
  & 68 & $^{95}$Mo & 8   & 8  &     &     \\
  & 69 & $^{94}$Mo & 8   & 8  &     &     \\
  & 70 & $^{92}$Mo & 8   & 8  &     &     \\
  & 71 & $^{nat}$Mo& 113 & 38 &     &     \\
\hline 
54&    &            &     &    & 145 & 94  \\
  & 72 & $^{126}$Xe & 2   & 0  &     &     \\
  & 73 & $^{124}$Xe & 6   & 2  &     &     \\
  & 74 & $^{nat}$Xe & 1   & 1  &     &     \\
\hline 
55&    &            &     &    & 370 & 370 \\
  & 75 & $^{133}$Cs & 83  & 83 &     &     \\
\hline 
56&    &            &     &    & 834 & 238 \\
  & 76 & $^{nat}$Ba & 79  & 30 &     &     \\
\hline 
57&    &            &     &    & 122 & 122 \\
  & 77 & $^{nat}$La & 66  & 66 &     &     \\
\hline 
77&    &            &     &    & 114 & 94  \\
  & 78 & $^{93}$Ir  & 1   & 1  &     &     \\
  & 79 & $^{nat}$Ir & 46  & 28 &     &     \\
\hline 
79&    &            &     &    & 2104& 935 \\
  & 80 & $^{197}$Au & 278 &271 &     &     \\
\hline 
80&    &            &     &    & 73  & 73  \\
  & 81 & $^{202}$Hg & 10  &10  &     &     \\
\hline 
83&    &            &     &    & 1142& 995 \\
  & 82 & $^{209}$Bi & 262 &174 &     &     \\
\hline 
\hline 
\\
{\bf Total:} &{\bf 82} &    &{\bf 2272} &{\bf 1574}&{\bf 22679} & {\bf 13389}\\
\end{tabular}
\end{center}

\newpage
\noindent{
the format and 
of references. Our library is still in progress and we hope to extend
it, depending on our needs, and to make it available for users through
the Web.  }

Presently, our experimental library contains 22,679 data points for 82 targets
of 32 elements covering 2,272 proton-induced reactions.
We also have begun to store in our library data for 
intermediate energy neutron-induced
reactions, but so far we have only 95 data points for Bi and C
targets covering 14 reactions induced by neutrons.

For comparison, we also show in Table 1 the statistics of available
data for the same targets in NUCLEX~\cite{nuclex}.
One can see, that for several targets like Cl, K, Ga, Ge, Cs, La, and
Hg, no new measurements were performed in recent years, and we 
have not found more data than is in NUCLEX. On the other hand, for such
targets like Ca, Fe, Co, Y, Zr, Ba, and Au, we have available
2-3 times more data points than in NUCLEX. Also, there are a number
of targets like $^{12}$C, $^{14}$N, $^{16}$O, $^{20}$Ne, $^{32}$S, $^{40}$Ar, 
$^{40}$Ca, and $^{126}$Xe, for which there are no data at all in NUCLEX,
while we presently have data for 181 reactions on these targets.

\begin{center}
{\large 3. Calculated Cross Section Library} \\
\end{center}

We needed to calculate a library of isotope production 
by nucleons at energies above 100 MeV
to simulate the production of medical radioisotopes in a high-energy
neutron and proton environment, e.g., at an APT Facility
\cite{medical97,medical98}, to benchmark \cite{report97}
the CEM95 code \cite{cem95}, and to see how a similar library created
by M. B. Chadwick \cite{mark} using the HMS-ALICE code \cite{alice96}
for energies below 150 MeV agrees with calculations at higher 
energies \cite{act150}. 

We use the CEM95 code \cite{cem95} to perform most of the calculations
for our activation library above 100 MeV. A number of 
reactions are calculated also
with the LAHET code system (version 2.83) \cite{lahet,lahet283},
and just a few reactions, with the recently improved version of the CEM code
\cite{cem98}. For energies below 100 MeV, we use the LA150
activation library by M. B. Chadwick \cite{mark}.

We store calculated cross sections for the production of all possible isotopes 
from 78 targets (see the list in Table 2), and then use the results we need.
Most of the reactions are calculated for energies from 100 MeV to 1.7 GeV,
according to the needs of our medical isotope production study 
\cite{medical97,medical98}, while several reactions 
are calculated only up to 1.0 GeV for \cite{mike98}, while some others used
extensively in our benchmark \cite{report97,act150}, were
calculated from 10 MeV to 5 GeV. Having already a tested method for
production of activation cross sections at these energies, our
library can be extended easily for other targets, when necessary.

We have benchmarked our calculations against all available experimental
data. Figures with comparison of more than 1000 excitation functions
calculated with CEM95, LAHET, and HMS-ALICE codes with the
available experimental data and predictions of other well-known models
can be found in \cite{report97,act150,medical97,medical98}. From this comparison
we see that, for the majority of nuclides
the results agree quite well with each other and with the data.
We can therefore conclude that the calculated activation cross sections
are reasonably reliable and can be used together with available
experimental data to produce an evaluated activation library, which we
describe in the next section.

\begin{center}
{\large 4. Evaluated Library} \\
\end{center}

As we mentioned previously, neither available experimental data nor any
of the current models or phenomenological systematics
can be used alone to produce a reliable evaluated activation
library covering a large area of target nuclides and incident
energies. Therefore, we choose to create our evaluated library 
by approximating by hand smoothly, wherever possible, excitation
functions using all available experimental data along with calculations 
using some more reliable codes, employing each of them in the 
corresponding regions of
targets and incident energies where they work better.
When we have reliable experimental data, they are taken as
the highest priority for our approximation as compared to model results.

\begin{center}
{\bf Table 2} \\
List of targets covered by the
CEM95 Activation Library: p(n) + A $\to$ any isotope\\
 ( $10/100$ MeV $\leq T_0 \leq 1.7/5.0$ GeV)
\end{center}
%\begin{sf}
\begin{tabbing}
123456789012345678901234567890123456789 \= 12345678901234567890123456789  \kill
$^{12}$C 
\>  
$^{58}$Ni, $^{60}$Ni, $^{61}$Ni, $^{62}$Ni,  $^{64}$Ni, $^{nat}$Ni \\
$^{14}$N
\>  
$^{63}$Cu, $^{65}$Cu, $^{nat}$Cu \\
$^{16}$O, $^{18}$O
\>  
$^{67}$Zn, $^{68}$Zn, $^{70}$Zn \\
$^{19}$F
\>  
$^{69}$Ga, $^{71}$Ga \\
$^{21}$Ne, $^{22}$Ne
\>  
$^{70}$Ge, $^{73}$Ge, $^{74}$Ge, $^{76}$Ge \\
$^{23}$Na
\>  
$^{75}$As \\
$^{24}$Mg, $^{25}$Mg, $^{26}$Mg , $^{nat}$Mg
\>  
$^{89}$Y \\
$^{27}$Al
\>  
$^{90}$Zr, $^{91}$Zr, $^{92}$Zr, $^{94}$Zr, $^{96}$Zr, $^{nat}$Zr  \\
$^{32}$S, $^{33}$S, $^{36}$S , $^{nat}$S 
\>  
$^{93}$Nb \\
$^{35}$Cl, $^{37}$Cl
\>  
$^{92}$Mo, $^{94}$Mo, $^{95}$Mo, $^{96}$Mo, $^{97}$Mo, $^{98}$Mo, $^{100}$Mo\\
$^{36}$Ar, $^{38}$Ar, $^{40}$Ar
\>  
$^{132}$Xe, $^{134}$Xe \\
$^{39}$K, $^{40}$K, $^{41}$K 
\>  
$^{133}$Cs \\
$^{40}$Ca
\>  
$^{134}$Ba, $^{135}$Ba, $^{136}$Ba, $^{137}$Ba, $^{138}$Ba , $^{nat}$Ba \\
$^{54}$Fe, $^{56}$Fe, $^{57}$Fe, $^{58}$Fe, $^{nat}$Fe
\>  
$^{138}$La, $^{139}$La \\
$^{59}$Co
\>  
$^{197}$Au 
\end{tabbing}
%\end{sf}
			   	
The recent {\em International Code Comparisons for Intermediate Energy Nuclear
Data} organized by NEA/OECD at Paris \cite{paris94,paris97},
our own comprehensive benchmarks \cite{report97,act150,medical97,medical98},
and several studies by Titarenko et al. \cite{titarenko} have shown
that CEM95 and LAHET generally have the best predictive powers for
spallation reactions at energies above 100 MeV as compared to other
available models. Therefore, we choose them above 100 MeV to create our
evaluated library. Actually, we employ the calculated library described
in Sec. 3.
The same benchmarks have shown that at lower energies, the
HMS-ALICE code does one of the best jobs in comparison with other models.
So, we use the activation library calculated by M. Chadwick \cite{mark} 
with the HMS-ALICE code \cite{alice96} for protons
below 100 MeV and neutrons between 20 and 100 MeV. In the overlapping
region, between 100 and 150 MeV, we use both HMS-ALICE and CEM95 and/or
LAHET results. For neutrons below 20 MeV, data of the European
Activation File EAF-97, Rev. 1 \cite{eaf97} with some recent improvements by
M. Herman \cite{herman} seem to be the most reliable, therefore we use 
them as the first priority in our evaluation.

Measured cross-section data from the compilation described in Sec. 2,
when available, are included together with theoretical results
and are used
to evaluate cross sections for our medical isotope production
study \cite{medical97,medical98} for 70 nuclides of 25 elements.
We note that when we put together all these different theoretical
results and experimental data, rarely do they agree perfectly with each 
other, providing a smooth continuity of evaluated excitation functions. 
Often, the resulting compilations show significant disagreement at energies 
where the available data progresses from one source to another. These sets are
thinned to eliminate discrepant data, providing data sets of more-or-less 
reasonable continuity defining our evaluated cross sections used in 
calculations \cite{medical97,medical98}.

Examples with typical results of evaluated activation cross sections 
for several proton and neutron reactions are shown in Figs. 1 and 2
by broad gray lines. 51 similar color
figures for proton-induced reactions and 56 figures for neutrons,
can be found on the Web, in our detailed report \cite{medical98}.
\newpage

\begin{figure}[h!]
%\vspace*{+4.0cm}
%\centerline{
%\psfig{figure=figures/fer0.ps,width=140mm,angle=0}\hspace{-0mm}}
%\psfig{figure=p1.ps,width=150mm,angle=0}\hspace{-50mm}}
\hspace*{10mm}
\psfig{figure=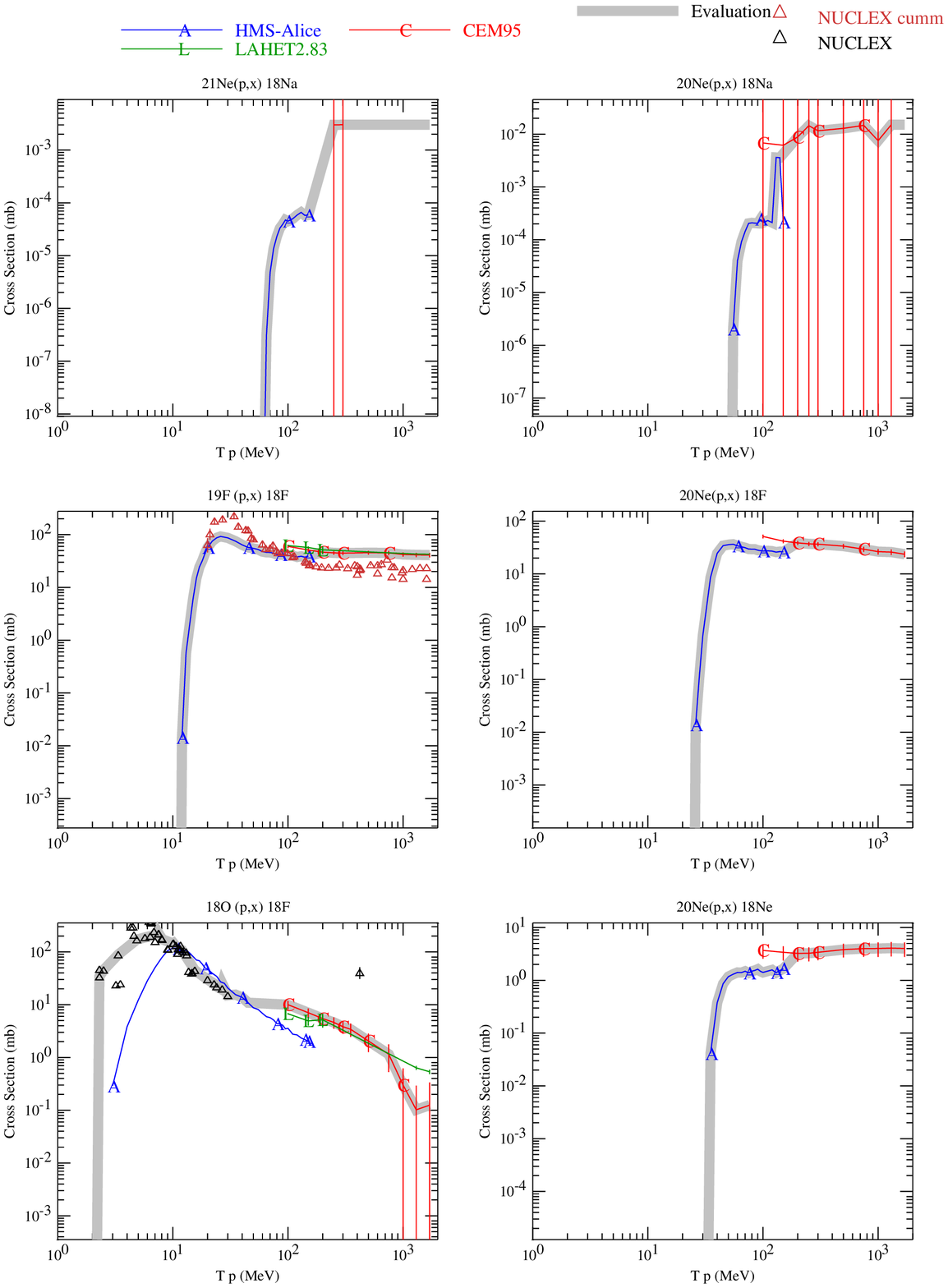,width=150mm,angle=0}
%}
\end{figure}
%\vspace*{-0.8cm}

{\small
{\bf Figure~1}.
Example of several evaluated proton-induced activation cross
sections. Evaluated cross sections are shown by broad gray lines,
other notation are given in the plots and described in the text and
in \cite{medical98}.
}
\newpage

\begin{figure}[h!]
%\vspace*{+4.0cm}
%\centerline{
%\psfig{figure=figures/fer0.ps,width=140mm,angle=0}\hspace{-0mm}}
\hspace*{10mm}
\psfig{figure=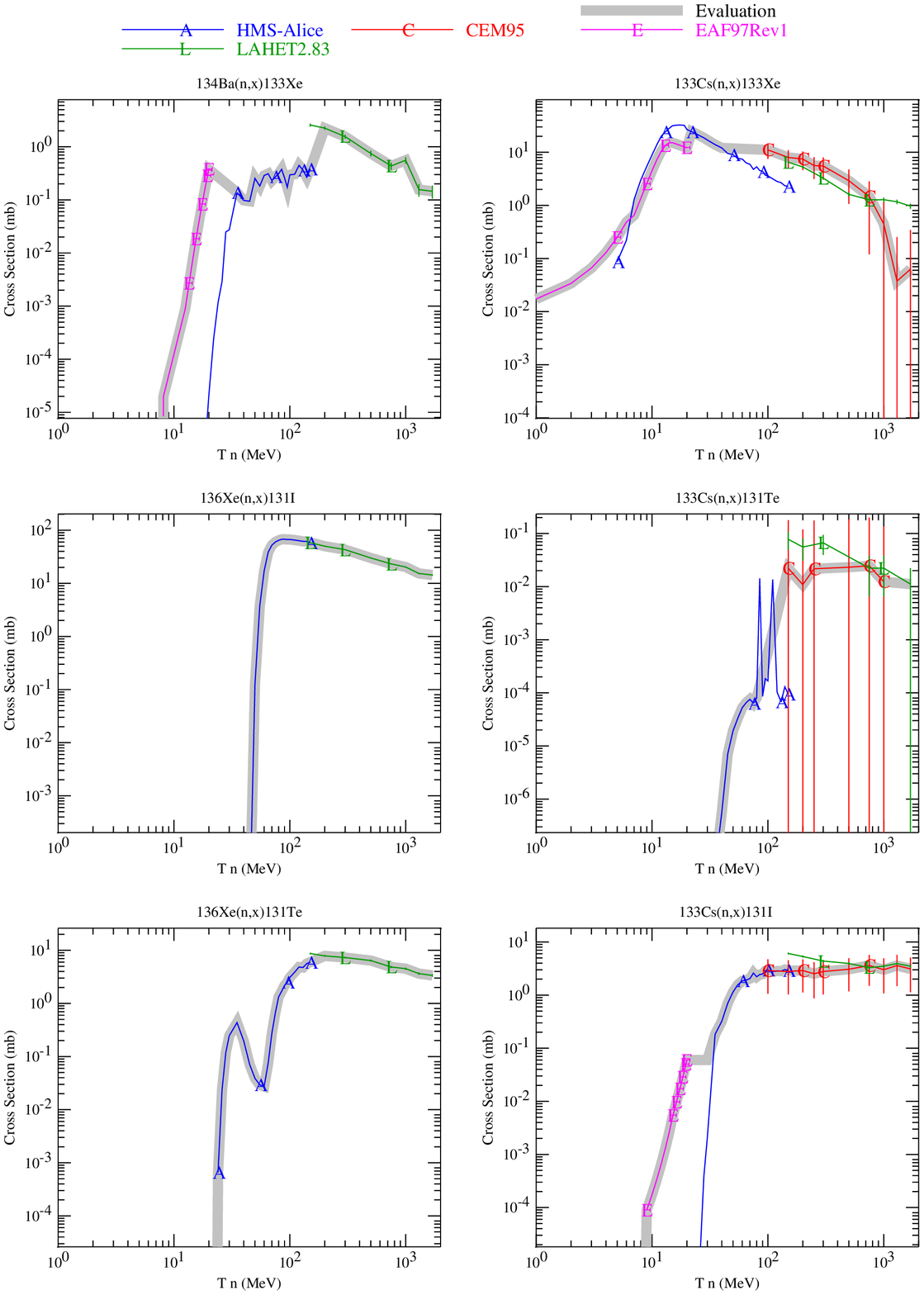,width=150mm,angle=0}
\end{figure}
%\vspace*{-0.5cm}

{\small
{\bf Figure~2}.
The same as in Fig. 1, but for neutron-induced reactions.\\
}

\newpage
\begin{center}
{\large 5. Summary} \\
\end{center}

We have produced experimental, calculated, and evaluated activation libraries 
for the interaction of nucleons with
nuclides covering about a third of all natural elements.
Our compilation of experimental proton-induced cross sections
contains 22,679 data points for 82 targets
of 32 elements covering 2,272 proton-induced reactions
and is the most complete we are aware of for these targets.

The methods developed are applicable to an extended set of reactions.
We plan to extend our libraries, depending on our needs, and to make 
them available for users through the Web. \\
\begin{center}
{\it Acknowledgements } 
\end{center}
We express our gratitude to  R.~E.~MacFarlane and 
L.~S.~Waters for interest in and support of the present work.
This study was supported by the U.~S.~Department of Energy.

%\vspace*{-1.0cm}


\begin{thebibliography}{99}

\vspace*{-0.1cm}
\bibitem{report97}
S. G. Mashnik, A. J. Sierk, O. Bersillon, and
T. A. Gabriel,
%``Cascade-Exciton Model Detailed Analysis of Proton Spallation at
%Energies From 10 MeV to 5 GeV,"
{\em Nucl. Instr. Meth.}, {\bf A414} (1998) 68;
LANL Report LA-UR-97-2905 (1997);
{\bf http://t2.lanl.gov/publications/publications.html}. 

\vspace*{-0.1cm}
\bibitem{cem98}
S.~G.~Mashnik and A.~J.~Sierk, 
``Improved Cascade-Exciton Model of Nuclear Reactions'', 
LANL Report LA-UR-98-5999 (1998);
this conference.
%Proc.~of the Fourth Workshop on Simulating Accelerator Radiation 
%Environments (SARE4), Knoxville, TN, September 14-16, 1998.

\vspace*{-0.1cm}
\bibitem{cem}
K. K. Gudima, S. G. Mashnik, and V. D. Toneev,
%``Cascade-Exciton Model of Nuclear Reactions,"
{\it Nucl. Phys.}, {\bf A401} (1983) 329.

\vspace*{-0.1cm}
\bibitem{cem95}
S. G. Mashnik,
{\em User Manual for the Code CEM95}, JINR, Dubna (1995),
OECD NEA Data Bank, 
%Le Saint-Germain 12,
%Boulevard des Iles, F-92130, Issy-les-Moulineax, 
Paris, France (1995);   
{\bf http://www.nea.fr/abs/html/iaea1247.html};
RSIC-PSR-357, Oak Ridge, 1995.

\vspace*{-0.1cm}
\bibitem{alice96}
M. Blann,
%``New Precompound Decay Model," 
{\em Phys. Rev.}, {\bf C54} (1996) 1341;
M. Blann and M. B. Chadwick,
%``New Precompound Decay Model: Angular Distributions,"
{\em Phys. Rev.}, {\bf C57} (1998) 233.

\vspace*{-0.1cm}
\bibitem{mark}
M. B. Chadwick, private communication on 150-MeV n \& p calculations, to
be published.                                                     

\vspace*{-0.1cm}
\bibitem{act150}
A. J. Koning, M. B. Chadwick, R. E. MacFarlane, S. G. Mashnik, and 
W. B. Wilson,
``Neutron and Proton Transmutation/Activation Libraries Up to 150 MeV,"
to be published.

\vspace*{-0.1cm}
\bibitem{medical97}
K. A. Van Riper,
S. G. Mashnik, M. B. Chadwick, M. Herman, A. J. Koning,
E. J. Pitcher, A. J. Sierk, G. J. Van Tuyle, L. S. Waters, and
W. B. Wilson,
%``APT Medical Isotope Production Study: $^{18}$F and $^{131}$I Production,"
LANL Report LA-UR-97-5068 (1997);
{\bf http://t2.lanl.gov/publications/publications.html}.

\vspace*{-0.1cm}
\bibitem{medical98}
K. A. Van Riper,
S. G. Mashnik,  and
W. B. Wilson,
``Study of Isotope Production in High Power Accelerator,"
this conference and LANL Report LA-UR-98-5378 (1998); see
detailed report in
LANL Report LA-UR-98-5379 (1998)
and at:
{\bf http://t2.lanl.gov/publications/publications.html}.

\vspace*{-0.1cm}
\bibitem{mike98}
M. James, S. Maloy, W. Sommer, M. Fowler, G. Mueller, and K. Corzine,
``Proton/Neutron Fluences on Mechanical Property Samples for the 
Accelerator Production of Tritium Project,"
this conference. 


\vspace*{-0.1cm}
\bibitem{nuclex} 
A.~S.~Iljinov, V.~G.~Semenov, M.~P.~Semenova, N.~M.~Sobolevsky,
and L.~V.~Udovenko,
%``Production of Radionuclides at Intermediate Energies,"
Springer Verlag, Landolt-B\"ornstein, New Series, 
subvolumes {\bf I/13a} (1991), {\bf I/13b} (1992), {\bf I/13c} (1993), 
{\bf I/13d} (1994), 
{\bf I/13e} (1994), {\bf I/13f} (1995), {\bf I/13h} (1996), 
{\bf I/13h} (1996);
V.~I.~Ivanov, N.~M.~Sobolevsky, and V.~G.~Semenov,
``NUCLEX --- an IBM PC Version of Handbook on
Radionuclide Production Cross Section at Intermediate Energies,"
{\em Proc.~Specialists' Mtg.}, Issy-les-Moulineaux, France, 
May 30--June 1, 
1994, OECD, p.~387;
%pp.~387--388 
``Computer Version Of the Handbook on Radionuclide Production
Cross-Sections at Intermediate Energies (the NUCLEX Code),"
{\em Proc. 3d Specialists Meeting on Shielding Aspects of Accelerators,
Targets and Irradiation Facilities (SATIF-3)}, Tohoku University,
Sendai, Japan, May 12-13, 1997, NEA/OECD (1998) 
p. 277.

\vspace*{-0.1cm}
\bibitem{michel}
R. Michel et. al.,
%R.~MICHEL, R.~BODEMANN, H.~BUSEMANN, R.~DAUNKE, M.~GLORIS, H.-J.~LANGE,
%B.~KLUG, A.~KRINS, I.~LEYA, M.~L\"UPKE, S.~NEUMANN, H.~REINHARDT,
%M.~SCHNATZ-B\"UTTGEN, U.~HERPERS, Th.~SCHIENKEL, F.~SUDBROCK, B.~HOLMQVIST,
%H.~COND\'E, P.~MALMBORG, M.~SUTER, B.~DITTRICH-HANNEN, P.-W.~KUBIK,
%H.-A.~SYNAL, and D.~FILGES
%``Cross Sections for the Production of Residual Nuclides by Low- and
%Medium-Energy Protons from the Target Elements C, N, O, Mg, Al, Si, Ca, Ti, V, 
%Mn, Fe, Co, Ni, Cu, Sr, Y, Zr, Nb, Ba, and Au,"
{\em Nucl. Instr.~Meth.}, {\bf B129} (1997) 153;
% E. Gilabert, B. Lavielle, S. Neumann, M. Gloris, R. Michel,
% Th. Schiekel, F. Sudbrock, and U. Herpers, " Cross Sections for the
% Proton-Induced Production of Krypton Isotopes from Rb, Sr, Y, and Zr
% for Energies up to 1600 MeV,"
{\em ibidem}, {\bf B145} (1998) 293
% pp. 293-319 
and references
therein;
% No. 2, pp. 153-193
see also the Web page at:\\
{\bf http://sun1.rrzn-user.uni-hannover.de/zsr/survey.htm\#url=overview.htm}.

\vspace*{-0.1cm}
\bibitem{titarenko}
Yu. E. Titarenko et al.,
%, O. V. Shvedov, M. M. Igumnov, R. Michel, S. G. Mashnik,
%E. I. Karpikhin, V. D. Kazaritsky, V. F. Batyaev, A. B. Koldobsky,
%V. M. Zhivun, A. N. Sosnin, R. E. Prael, M. B. Chadwick,
%T. A. Gabriel, and M. Blann,
%``Experimental and Theoretical Study of the Yields of
%Radionuclides Produced in $^{209}$Bi Thin Targets Irradiated by
%130 MeV and 1.5 GeV Protons,"
{\em Nucl. Instr. Meth.}, {\bf A414} (1998) 73;
%Los Alamos National Laboratory Report LA-UR-97-3787 (1997);
{\em E-print:} {\bf nucl-th/9709056};
%Yu. E. Titarenko, O. V. Shvedov, V. F. Batyaev, E. I. Karpikhin, 
%V. M. Zhivun, A. B. Koldobsky, M. M. Igumnov, I. S. Sklokin, 
%R. D. Mulambetov, A. N. Sosnin, H. Yasuda, H. Takada, S. Chiba, Y. Kasugai,
%S. G. Mashnik, R.E. Prael, M.B. Chadwick, T.A. Gabriel, and M. Blann,
%``Experimental and Computer Simulation Study of  
%Radionuclide Formation in the ADT Materials Irradiated with 
%Intermediate Energy Protons,"
{\em Proc. Second Int. Topical Meeting on Nuclear Applications of Accelerator 
Technology (AccApp'98)}, Gatlinburg, TN, USA, September 20-23, 1998 
and references therein. 

\vspace*{-0.1cm}
\bibitem{al94}
Yu. V. Aleksandrov et al.,
%Yu.~V.~ALEKSANDROV, S.~K.~VASILJEV, R.~B.~IVANOV, L.~M.~KRIZHANSKY,
%M.~A.~MIKHAILOVA, T.~I.~POPOVA, V.~P.~PRIKHODTZEVA, V.~P.~EISMONT,
%A.~F.~NOVGORODOV, and R.~MISIAK,
%``Cross Section for Spallation Production of Radioactive
%Nuclides by 660 MeV Protons," 
{\em Izv.~Rossiiskoi Akad.~Nauk, ser.~fiz.}, {\bf 59} (1995) 206
%  No.5, pp.206--210
[{\em Bull.~Russian Acad.~Sci.: Physics}, {\bf 59} (1996) 895]
and references therein.
% No.~5, pp.~895--899

\vspace*{-0.1cm}
\bibitem{bel92}  
B. N. Belyaev, V. D. Domkin, and V. S. Mukhin, 
%``Isotopic Effects in Spallation and Fission Reactions and 
%Mass-Spectrometer Investigation of Them with 1-GeV Protons," 
{\em Fiz. Elem. Chastits At. Yadra}, {\bf 23} (1992) 993
[{\em Sov. J. Part. Nucl}., {\bf 23} (1992) 439];
%``Isotope Effects in Fragment Yields in Uranium Nuclear Fission 
%Induced by Intermediate-Energy Protons,
{\em Yad. Fiz.}, {\bf 57} (1994) 1231 
[{\em Phys. At. Nucl.}, {\bf 57} (1994) 1163].

\vspace*{-0.1cm}
\bibitem{ven93}
 N. I. Venikov, V. I. Novikov, and A. A. Sebiakin, 
% "Excitation Functions
%      of Proton-induced Reactions on 126-Xe: 125-I Impurity in 123-I,"
{\em Appl. Radiat. Isot.}, {\bf 44} (1993) 751. 

\vspace*{-0.1cm}
\bibitem{dan97}
A. S. Danagulyan et al., 
%L. G. Martirosyan, N. S. Amelin, A. R. Balabekyan,
%V. G. Kalinnikov, V. I. Stegailov, and Y. Frana,
%``Investigation of (p,xn), (p,pxn), and (p,2p) Reactions on Tin Isotopes,"
{\em Yad. Fiz.}, {\bf 60} (1997) 965.

\vspace*{-0.1cm}
\bibitem{von97}
H. Vonach et al., 
%A. Pavlik, A. Wallner, M. Drosg, R. C. Haight, D. M. Drake, and S. Chiba, 
%"Spallation Reactions in Al-27 and Fe-56 Induced by 800 MeV Protons," 
{\em Phys. Rev.}, {\bf C55} (1997) 2458.

\vspace*{-0.1cm}
\bibitem{su94}
S. Sudar and S. M. Qaim, 
%"Excitation Functions of Proton and Deutron
%       Induced Reactions on Iron and Alpha-Particle Induced Reactions on 
%       Manganese in the Energy Region up to 25 MeV," 
{\em Phys. Rev.}, {\bf C50} (1994) 2408.

\vspace*{-0.1cm}
\bibitem{ba97}
D. W. Bardayan, 
%M. T. F. da Cruz, M. M. Hindi, A. F. Barghouty, Y. D. Chan, 
%A. Garcia, R.-M. Larimer, K. T. Lesko, E. B. Norman, D. F. Rossi, 
%F. E. Wietfeldt, and I. Zlimen, 
%      "Radioactive Yields from 1.85-GeV Protons on Mo and 1.85- and 5.0-GeV 
% Protons on Te,"
{\em Phys. Rev.}, {\bf C55} (1997) 820. 

\vspace*{-0.1cm}
\bibitem{si97}
J. M. Sisterson et al., 
%K. Kim, A. Beverding, P. A. J. Englert, M. Caffee, A. J. T. Jull, 
%D. J. Donahue, L. McHargue, J. Vincent, and R. C. Reedy,
%       "Measurement of Proton Production Cross Sections of 10-Be and 26-Al
%       from Elements Found in Lunar Rocks,"
{\em Nucl. Instr. Meth.}, {\bf B123} (1997) 324.

\vspace*{-0.1cm}
\bibitem{web90}
W. R. Webber, J. C. Kish, and D. A. Schrier, 
%"Individual Isotopic
%       Fragmentation Cross Sections of Relativistic Nuclei in Hydrogen,
%       Helium, and Carbon Targets," 
{\em Phys. Rev.}, {\bf C41} (1990) 547.

\vspace*{-0.1cm}
\bibitem{chen97}
C.-Z. Chen et al.,
%, S. Albergo, Z. Caccia et al., 
%"Systematics of Isotopic 
%       Production Cross Sections from Interactions of Relativistic Ca-40
%       in Hydrogen," 
{\em Phys. Rev.}, {\bf C56} (1997) 1536.

\vspace*{-0.1cm}
\bibitem{ta98}
L. Tassan-Got et al., 
%B. Mustapha, F. Farget, M. Bernas, C. Stephan, P. Armbruster, J. Benlliure, 
%T. Enqvist, K. H. Schmidt, A. Boudard, R. Legrain, S. Leray, C. Volant, 
%W. Wlazlo, S. Czajkowski, and M. Pravikoff, 
"Spallation Residue Cross-Sections in Reverse  Kinematics," 
Proc. Int. Conf. on the Phys. of Nucl. Sci. and Techn.,
 October 5-8, 1998, Long Island, Ney York, vol. 2, pp. 1334-1340.

\vspace*{-0.1cm}
\bibitem{lahet}
R.~E.~Prael and H.~Lichtenstein,
``User Guide to LCS: The LAHET Code System,"
{\em LA-UR-89-3014}, LANL (September 1989).
%pp. 277-286.

\vspace*{-0.1cm}
\bibitem{lahet283}
R. E. Prael and D. G. Madland, 
``LAHET Code System Modifications for {\bf LAHET 2.8},"
{\em LA-UR-95-3605}, LANL (September 1996).

\vspace*{-0.1cm}
\bibitem{paris94}
M. Blann, H. Gruppelar, P. Nagel, and J. Rodens,
{\em International Code Comparison for Intermediate Energy Nuclear Data},
NEA OECD, Paris (1994).

\vspace*{-0.1cm}
\bibitem{paris97}
R. Michel and P. Nagel,
{\it International Codes and
Model Intercomparison for Intermediate Energy Activation Yields},
NSC/DOC(97)-1, 
%NEA/P/{\&}T No 14, 
OECD, Paris (1997);  
{\bf http://www.nea.fr/html/science/pt/ieay}.
 

\vspace*{-0.1cm}
\bibitem{eaf97}
J.-Ch. Sublet, J. Kopecky, R. A. Forrest, and D. Nierop,
{\em The European Activation File: EAF-97 Report file-Rev. 1},
UKAEA, Culham, Abigdon, Oxfordshire OX 14 3DB, United Kingdom
(December, 1997). 

\vspace*{-0.1cm}
\bibitem{herman}
M. Herman,
{\em LANL Update II of the ECNAF Neutron Activation Cross-Section Library},
Los Alamos National Laboratory Report LA-UR-96-4914 (December, 1996).
\end{thebibliography}
\end{document}